\documentclass[11pt,a4paper]{amsart}
\usepackage{amssymb,amsthm,euscript,amsmath,graphicx,array}

\newcommand{\be}{\begin{equation}}
\newcommand{\ee}{\end{equation}}
\newcommand{\ben}{\begin{equation*}}
\newcommand{\een}{\end{equation*}}
\newcommand{\bea}{\begin{eqnarray}}
\newcommand{\eea}{\end{eqnarray}}
\newcommand{\bean}{\begin{eqnarray*}}
\newcommand{\eean}{\end{eqnarray*}}
\newcommand\re[1]{(\ref{#1})}

\newcommand{\eqnsection}{
\renewcommand{\theequation}{{\thesection.\arabic{equation}}}
\makeatletter
\csname @addtoreset\endcsname{equation}{section}
\makeatother}
\eqnsection

\newlength{\picwidth}
\newcommand{\fig}[4]
{\begin{\center}
 \includegraphics[width=#4]{pictures//#2}\\[2mm]
 Fig.~#1:~#3
\end{center}}
\begin{document}


\title{The effects of nonlocality on the evolution of higher order
fluxes in non-equilibrium thermodynamics} \maketitle \centerline{V.
A. Cimmelli and P. V\'an\footnote{Corresponding author.} }

\date{}

\begin{abstract}
The role of gradient dependent constitutive spaces is investigated
on the example of Extended Thermodynamics of rigid heat
conductors. Different levels of nonlocality are developed and the
different versions of extended thermodynamics are classified. The
local form of the entropy density plays a crucial role in the
investigations. The entropy inequality is solved under suitable
constitutive assumptions. Balance form of evolution equations is
obtained in special cases. Closure relations are derived on a
phenomenological level.
\medskip
\medskip
\medskip
\medskip

\medskip
\medskip


\end{abstract}

\markboth{\small\scshape V. A. Cimmelli and P.
V\'an}{\small\scshape Nonlocal effects in non-equilibrium
thermodynamics}
\date{}

\section{Introduction}

Weakly nonlocal thermodynamic theories are those that introduce
the space derivatives of the basic variables into constitutive
functions \cite{Van03a}. Second Law restricts considerably the
form of the constitutive quantities and gives a genuine insight
into the structure of the theories. Weakly nonlocal constitutive
functions are mostly investigated in relation of material
microstructure in mechanics \cite{Val98a,Mar02a} or to find
nonlocal extensions of classical theories \cite{Van03a}.

In this paper we investigate nonlocal constitutive spaces with
different levels of nonlocality, namely of different order of
derivatives. However, we derive also the restrictions that are due
to locality assumptions on different levels. In our analysis we
assume a non-equilibrium entropy function that can be approximated
by its values measured at the equilibrium. Such an assumption is
referred to as {\em local state} hypothesis \cite{Kes93a1}.

We restrict ourselves to extended thermodynamic theories of rigid
heat conductors \cite{Ver97b,JouAta92b,MulRug98b} and introduce
the heat flux together with a second order tensor as internal
variables. The balance structure of the evolution equations is not
postulated. Furthermore, the entropy current is regarded as a
constitutive quantity and we are to give a complete solution of
the thermodynamic constraints i.e. both the equalities and the
residual dissipation inequality.

In a previous work \cite{CiaAta03m} the local theory has been
developed in the details and the evolution equations for fluxes of
higher tensorial order have been obtained. Also it was proved that
under particular assumptions on the entropy density and the
entropy current the balance form can be recovered. Moreover, the
system of equations was closed, in that the evolution equations
for the highest order variable in the hierarchy - ordinary
differential equations - can be interpreted as a closure relation.

In the present paper we extend our investigation to the case of
weakly nonlocal constitutive state spaces. The solutions are
derived with the help of minimal assumptions on the form of the
entropy, on its flux or on the evolution equations of the internal
variables. In this way the different solutions of the entropy
inequality are clearly classified. The most general assumption,
that covers all existing phenomenological theories lies on the
concept of {\em current multipliers}, which represent some
constitutive functions entering the entropy current. We will see
that the final evolution equations are more general than the
traditional balance ones. The conditions to recover the classical
cases are clarified.

In Section 2 we investigate first order nonlocality by applying
Liu procedure \cite{Liu72a,HauKir02a} for the exploitation of
Second Law.

In Section 3 we solve the Liu equation in the case of local state
and local evolution equations for the dynamic variables.  These
assumptions result in a set of rather unusual restrictions from
which we argue that some kind of nonlocality, either of the
constitutive space or of the evolution equations, seems to be
unavoidable. On the other hand, if we face with nonlocal state
but local evolution equations then the entropy current is local,
provided that the entropy density does not depend on the gradient
of the internal energy.

In Section 4 we investigate the traditional assumptions of
Extended Irreversible Thermodynamics based on the following form
of the local entropy \cite{Gya77a}
\be
\label{Gyarent} s(e,q_i,\Phi_{ij}) = s_0 -\frac{1}{2} m_{ij}q_i
q_j - \frac{1}{2} n_{ijkl} \Phi_{ij} \Phi_{kl},
\ee

\noindent where the matrices $m_{ij}$ and $n_{ijkl}$ are
constitutive functions and $s_0$ is the {\em equilibrium entropy},
that depends only on the internal energy.  We first suppose the
entropy current is given as \cite{Nyi91a1}
\be j_i = A_{ij}q_j + B_{ijk}\Phi_{jk}, \label{Nyiecurr}\ee
\noindent where $A_{ij}$ and $B_{ijk}$ are constitutive functions,
the so called {\em current multipliers}. Then we explore the less
general case \cite{Ver83a}, too
\be\label{Verecurr} j_i = \frac{\partial s}{\partial e} q_i +
  \frac{\partial s}{\partial q_k} \Phi_{ki}
\ee
It is worth noticing that the general form of the entropy current
\re{Nyiecurr} reduces to \re{Verecurr} when $A_{ij} =
\frac{\partial s}{\partial e} \delta_{ij}$ and $B_{ijk} =
\frac{\partial s}{\partial q_j} \delta_{ik}$. We investigate
different assumptions that can be compatible with the balance form
of the evolution equations.

In Section 5 we consider second order nonlocality but conserve
the form \re{Gyarent} of the entropy density and the expression
\re{Nyiecurr} of the entropy current. In such a case, due to the
enlargement of the state space, the balance form can be obtained
even if the general constitutive equation \re{Nyiecurr} holds
true. We show that all previous examples can be recovered under
simple special assumptions.

In Section 6 we point out some nonlocal effects arising in
thermal wave propagation  at low temperature, which are described
by the celebrated Guyer-Krumhansl equation
\cite{GuyKru66a1,GuyKru66a2,AckGuy68a}. Such an equation has been
derived by the authors by solving a linearized Boltzmann equation
for phonon gas hydrodynamic. Here we prove that it can be obtained
in the classical macroscopic framework of nonlocal irreversible
thermodynamics.

The previous results are discussed in Section 7, where a table
shows the connections between the constitutive assumptions and the
thermodynamic restrictions, as far as the locality and
nonlocality are concerned.

\section{First order nonlocality - exploitation of the Second Law}

In a rigid heat conductor at rest we start from the following local
balance equation of the internal energy
\be
  \dot{e} + q_{i'i} =0, \label{balinte}
\ee \noindent where $e$ is the density of internal energy, $q_i$
$i=1,2,3$ are the components of the heat flux, $\dot{f} \equiv
\frac{\partial f}{\partial t}$, $f_{'i} \equiv \frac{\partial
f}{\partial x_i}$, $x_i$ $i = 1,2,3$ are the Cartesian coordinates
of the points of the body and the Einstein convention of summation
over the repeated indices has been applied. The only equilibrium
variable will be the internal energy $e$, while the first dynamic
variable is supposed to be the heat flux ${q_i}$. As a further
dynamic variable let us choose a second order tensor, whose
components will be denoted by $\Phi_{ij}$, $i,j=1,2,3$. Hence the
{\em basic state space} (the wanted fields) in our investigations
will be spanned by the variables $(e,{q_i}, \Phi_{ij})$. This is a
13 field theory, because the number of the independent fields is
13. However, some reductions are possible. For instance in
Extended Thermodynamics tensor $\Phi_{ij}$ is identified with the
symmetric momentum flux $N_{[ij]}$ coming from kinetic theory
\cite{MulRug98b}. Then the unknown fields reduce to 10 and we face
with a 10-field theory. Furthermore it is possible to decompose
$N_{[ij]}$ into an isotropic  part, which is related to the
internal energy, and a deviatoric part according to the equation
$N_{[ij]}=\frac{1}{3}e\delta_{[ij]} + N_{<ij>}$, where $N_{<ij>}$
is symmetric and traceless \cite{DreStr93a}. In such a particular
case the internal energy coincides with one of the six independent
components of $N_{[ij]}$ and we deal with a 9-field theory. We are
investigating a first order weak nonlocality in all variables of
the basic state, therefore the {\em constitutive space} is spanned
by the basic state and its spacial derivatives, that is the fields
$(e,q_i,\Phi_{ij},e_{'i},q_{i'j},\Phi_{ij'k})$. We assume that the
evolution equations for the heat current $q_i$ and for $\Phi_{ij}$
can be written in the following rather general form
\bea
\dot q_i  &=& g_i,    \label{ev1}\\
\dot \Phi_{ij} &=& f_{ij},\label{ev2}
\eea
\noindent where $g_i$ and $f_{ij}$ are constitutive functions.
With the assumption of first order nonlocality the spacial
derivatives of the above equations give further restrictions
\cite{Cim04a,Van05a1}.
\bea
\dot{e_{'i}} + q_{j'ji} &=& 0, \label{dbalinte}\\
\dot q_{i'j}  - g_{i,j} &=& 0,    \label{dev1}\\
\dot \Phi_{ij'k} - f_{ij'k} &=& 0.\label{dev2}
\eea
These equations are sometimes referred to as {\em prolonged forms}
of the evolution equations \re{balinte}, \re{ev1} and \re{ev2}.

The local balance of entropy is given by
\be
 \dot{s} + j_{i'i} =\sigma_s, \label{balent}
\ee
\noindent with $s$ standing for the entropy density,  $j_i$
$i=1,2,3$ for the components of the entropy current and
$\sigma_s$ for the density of entropy production. Second Law of
Thermodynamics forces $\sigma_s$ to be nonnegative.

In the following we will investigate the restrictions from the
inequality of the Second Law with the general assumption that both
the entropy and the entropy flux are constitutive quantities. The
method of the exploitation is given by the Liu procedure
\cite{Liu72a}. However, according to our calculations, in the
present case a generalized Coleman-Noll \cite{ColNol63a} procedure
gives identical results.

Let us introduce the Lagrange-Farkas multipliers \cite{Liu72a,
HauKir02a} $\Gamma^1$, $\Gamma^2_i$ and $\Gamma^3_{ij}$ for the
evolution equations \re{balinte}, \re{ev1} and \re{ev2}
respectively. The multipliers $\Gamma^4_{i}$, $\Gamma^5_{ij}$ and
$\Gamma^6_{ijk}$ are for the prolonged evolution equations
\re{dbalinte}, \re{dev1} and \re{dev2} respectively.

Now, Liu procedure gives
\bean
\partial_1s\dot{e} &+&
    (\partial_2s)_i\dot{q}_{i} +
    (\partial_3s)_{ij}\dot{\Phi}_{ij} +
    (\partial_4s)_i\dot{e}_{'i} +
    (\partial_5s)_{ij}\dot{q}_{i'j} + \\
+ (\partial_6s)_{ijk}\dot{\Phi}_{ij'k} &+&
    (\partial_1 j_i)e_{'i} +
    (\partial_2 j_i)_j q_{j'i} +
    (\partial_3 j_i)_{jk} \Phi_{jk'i} +
    (\partial_4 j_i)_j e_{'ij} + \\
+ (\partial_5 j_i)_{jk} q_{j'ki} &+&
    (\partial_6 j_i)_{jkl} \Phi_{jk'li} - \\
- \Gamma^1\left(\dot{e} + q_{i'i}\right) &-&
    \Gamma^2_i \left(\dot{q}_i  - g_i\right) -
    \Gamma^3_{ij} \left(\dot{\Phi}_{ij}  - f_{ij}\right) -
    \Gamma^4_i \left(\dot{e}_{'i} + q_{j'ji} \right) -\\
- \Gamma^5_{ij} \left(\dot{q}_{i'j} \right. &-&
    (\partial_1 g_i)e_{'j} -
    (\partial_2 g_i)_k q_{k'j} -
    (\partial_3 g_i)_{kl} \Phi_{kl'j} -
    (\partial_4 g_i)_k e_{'kj} - \\
&-& \left. (\partial_5 g_i)_{kl} q_{k'lj} -
    (\partial_6 g_i)_{klm} \Phi_{kl'lmj}  \right) \\
- \Gamma^6_{ijk} \left(\dot{\Phi}_{ij'k}  \right. &-&
    (\partial_1 f_{ij})e_{'l} -
    (\partial_2 f_{ij})_l q_{l'k} -
    (\partial_3 f_{ij})_{lm} \Phi_{lm'k} -
    (\partial_4 f_{ij})_l e_{'lk} - \\
&-& \left. (\partial_5 f_{ij})_{lm} q_{l'mk} -
    (\partial_6 f_{ij})_{lmn} \Phi_{lm'nk}  \right) \geq 0.
\eean

Here $\partial_n$, $n = 1,2,3,4,5,6$ denotes the partial
derivatives of the constitutive functions according to the
variables $(e,q_i,\Phi_{ij},e_{'i},q_{i'j},\Phi_{ij'k})$
respectively. After some rearrangements of the inequality one
obtains the Lagrange-Farkas multipliers from the first set of the
Liu equations. These are obtained by imposing the coefficients of
the time derivatives vanish.
\bea
\partial_1s &=& \Gamma^1, \label{L1} \\
(\partial_2s)_{i} &=& \Gamma^2_{i}, \label{L2} \\
(\partial_3s)_{ij} &=& \Gamma^3_{ij}, \label{L3} \\
(\partial_4s)_{i} &=& \Gamma^4_{i}, \label{L4} \\
(\partial_5s)_{ij} &=& \Gamma^5_{ij}, \label{L5} \\
(\partial_6s)_{ijk} &=& \Gamma^6_{ijk}. \label{L6}
\eea

The second set of Liu equations is obtained by taking equal to
zero the multipliers of the second order space derivatives. By
applying \re{L1}-\re{L6} one can write them as
\bea
(\partial_4 j_i)_j +
    (\partial_5 s)_{il} (\partial_4 g_l)_j +
    (\partial_6 s)_{kli} (\partial_4 f_{kl})_j &=& 0, \label{L7} \\
(\partial_5 j_i)_{jk} -
    (\partial_4s)_i \delta_{jk} +
    (\partial_5 s)_{mi} (\partial_5 g_m)_{jk} +
    (\partial_6 s)_{mni} (\partial_5 f_{mn})_{jk}&=& 0, \label{L8} \\
(\partial_6 j_i)_{jkl} +
    (\partial_5 s)_{mi} (\partial_6 g_m)_{jkl} +
    (\partial_6 s)_{mni} (\partial_6 f_{mn})_{jkl}&=& 0. \label{L9}
\eea

Finally the residual dissipation inequality can be written in the
following form
\begin{eqnarray}
[\partial_1 j_i &+&
    (\partial_5 s)_{ji}\partial_1 g_j +
    (\partial_6 s)_{jki}\partial_1 f_{jk} ] e_{'i} + \nonumber\\
+[(\partial_2 j_j)_i &-&
    \partial_1 s \delta_{ij} +
    (\partial_5 s)_{kj}(\partial_2 g_k)_i +
    (\partial_6 s)_{klj}(\partial_2 f_{kl})_i ] q_{i'j} + \nonumber\\
+ [(\partial_3 j_k)_{ij} &+&
    (\partial_5 s)_{lk}(\partial_3 g_l)_{ij} +
    (\partial_6 s)_{lmk}(\partial_3 f_{lm})_{ij}
    ]\Phi_{ij'k} + \nonumber\\
&+& (\partial_2s)_i g_i + (\partial_3 s)_{ij} f_{ij} \geq 0.
\label{Dis1} \end{eqnarray}
It is easily seen that the Liu system  \re{L7}-\re{L9} is composed
by 117 differential equations constraining the set of the 832
partial derivatives of the constitutive functions $s, j_{i},
g_{i}, f_{ij}$ with respect to the elements of the constitutive
space $(e, q_i, \Phi_{ij}, e_{'i}, q_{i'j}, \Phi_{ij'k})$. Without
some simplifications there is no hope to solve such a system.

In the following sections we are looking for special simplifying
assumptions to solve the Liu equations \re{L7}-\re{L9} and the
dissipation inequality \re{Dis1}. First we will investigate cases
where some of the constitutive functions are assumed to be local.

\section{Solutions of the Liu equations - locality assumptions}

\subsection{Local state}

Let us start the assumption of the local state in the form that the
entropy is independent of the gradients:
\be
 s := {s}(e,q_i,\Phi_{ij}) \label{lstate}
\ee

In this case the Liu equations \re{L7}-\re{L9} are simplified
considerably and as a solution we obtain a local entropy current
\be
 j_i = {j_i}(e,q_i,\Phi_{ij}) \label{lcurr}.
\ee

The dissipation inequality \re{Dis1} simplifies, too
\begin{eqnarray}
\partial_1j_i e_{'i} &+&
    [(\partial_2 j_j)_i - \partial_1 s \delta_{ij}]q_{i'j} +
    (\partial_3 j_k)_{ij} \Phi_{ij'k} +
    (\partial_2s)_i g_i + (\partial_3 s)_{ij} f_{ij}  = \nonumber\\
&=& (j_i)_{'i} - \partial_1 s q_{i'i} +
    (\partial_2s)_i g_i + (\partial_3 s)_{ij} f_{ij}
    \geq 0.
\label{Dis1ls}\end{eqnarray}

\subsection{Local state and local evolution}

Let us investigate now the case when the evolution equations of the
internal variables are ordinary differential equations, that is we
assume that the constitutive quantities ${g_i}$ and ${f_{ij}}$ take
the local form:
\bea
{g_i} &:=& {g_i}(e,q_j,\Phi_{jk}) \label{lqev1} \\
{f_{ij}} &:=& {f_{ij}}(e,q_j,\Phi_{jk}) \label{lqev2}
\eea

Now the dissipation inequality simplifies further the possible
constitutive functions, because the coefficients of the
derivatives should disappear. These restrictions result in a
rather unusual material, since we get
\be
\label{simpl1} {j_i} = {j_i}({q_j}), \qquad \frac{\partial
j_i}{\partial q_j} =
    \frac{\partial s}{\partial e} \delta_{ij}.
\ee

On the other hand the dissipation inequality can be written in a
force-current form and can be solved for the constitutive
functions ${g_i}$ and $f_{ij}$
$$
(\partial_2s)_i g_i + (\partial_3 s)_{ij} f_{ij} \geq 0.
$$
By \re{simpl1} it follows that the temperature of the material
$\frac{\partial s}{\partial e}$ is independent of the internal
energy. Such a property is in contrast with the physical reality.
We conclude that some nonlocality is necessary in modelling rigid
heat conductors.

\subsection{Local evolution}

Let us assume now that the evolution of the internal variables is
local, but there is no local state, therefore we require
\re{lqev1} and \re{lqev2}, but \re{lstate} is not assumed. In this
case the entropy current is nonlocal, but the nonlocality is
rather reduced. The Liu equations \re{L7} and \re{L9} give that
the entropy current does not depend on the gradients of $e$ and
$\Phi$, and \re{L8} simplifies to
$$
\frac{\partial j_i}{\partial q_{j'k}} =
    \frac{\partial s}{\partial e_{'i}}\delta_{jk}
$$
The nonlocality in the ${q_i}$ is due to balance form of the
evolution equation of the internal energy \re{balinte}. From the
above constraint one can easily see, that the entropy current is
local if we further assume that the entropy is local in the
internal energy,
\ben
s := s(e,q_i,\Phi_{ij},q_{i'j},\Phi_{ij'k}).
\een

\section{Solutions of the entropy inequality in case of local state}

As we have seen above, the Liu equations are trivially solvable in
the local state. However, the solution of the dissipation
inequality can be achieved only with further assumptions.
Moreover, there are different assumptions to have physical models,
to introduce a suitable gradient dependencies. In Classical
Irreversible Thermodynamics \cite{Ver97b} and in Rational
Thermodynamics \cite{ColNol63a} the mentioned requirement of
nonlocality is achieved by introducing the gradient of temperature
(or, equivalently, of the internal energy) into the constitutive
space. In Extended Thermodynamics \cite{JouAta92b,MulRug98b} the
constitutive space is local but the evolution equations are
balances, they have a special nonlocal form. All kind of theories
of Extended Thermodynamics resulted in more of less satisfactory
models of different phenomena, but they provide different
solutions of the entropy inequality.

\subsection{Local state and special nonlocal evolution: linear nonlocality}

In this case the evolution equations depend linearly on the
gradients:
\bea
{g_i} &:=& A_{ij}e_{'j} + B_{ijk}q_{k'j} + C_{ijkl}\Phi_{kl'j}, \\
{f_{ij}} &:=& D_{ijk}e_{'k} + E_{ijkl}q_{l'k} + F_{ijklm}\Phi_{lm'k},
\eea

\noindent where $A_{ij}, B_{ijk}, C_{ijkl}, D_{ijk}, E_{ijkl},
F_{ijklm}$ are local constitutive functions.

Now the dissipation inequality \re{Dis1} reduces to a solvable
form as
\bea (\partial_1 j_i &+& (\partial_2 s)_j A_{jk} +
     (\partial_3 s)_{jk} D_{jki}  ) e_{'i} \\
+ ((\partial_2 j_j)_i &-& \partial_1s \delta_{ij}
    + (\partial_2 s)_k B_{kji}
    + (\partial_3 s)_{kl} E_{klji})q_{i'j} \\
+ ((\partial_3 j_k)_{ij} &+& (\partial_2 s)_l C_{lkij} +
    (\partial_3 s)_{lm} F_{lmkij} )\Phi_{ij'k} \geq 0.
\eea

As the quantities in the parentheses are local functions they
should be zero respectively. Therefore we get
\bea
\partial_1 j_i &=& - (\partial_2 s)_j A_{jk} -
     (\partial_3 s)_{jk} D_{jki}  \label{pcn1}\\
(\partial_2 j_j)_i &=&  \partial_1s \delta_{ij} - (\partial_2 s)_k B_{kji}
    - (\partial_3 s)_{kl} E_{klji} \label{pcn2}\\
(\partial_3 j_k)_{ij} &=& - (\partial_2 s)_l C_{lkij}
    - (\partial_3 s)_{lm} F_{lmkij}. \label{pcn3}
\eea

These equations cannot be solved without any further ado. However,
we can see that even if we do no know anything on the entropy
current $j_i$ they result in strong correlations on the entropy
derivatives and the evolution equation, as the mixed partial
derivatives of $j_i$ should be equal. On the other hand let us
observe that in this case the entropy production is zero, there is
no dissipation.

\subsection{Local state and special local evolution: balance form}

In this case one assumes, that the evolution equations have a
special balance form. Therefore there are potentials $Q_{ij}$  and
$H_{ijk}$ of the fields $(A_{ij}, B_{ijk}, C_{ijkl})(e, q_i,
\Phi_{ij})$ and $(D_{ijk}, E_{ijkl}, F_{ijklm})(e, q_i,
\Phi_{ij})$ respectively. The evolution equations can
be written as
\bea
{g_i} &:=& \partial_1 Q_{ij} e_{'j}
    + (\partial_2 Q_{ij})_k q_{k'j}
    + (\partial_3 Q_{ij})_{kl}\Phi_{kl'j}, \\
{f_{ij}} &:=&  \partial_1 H_{ijk} e_{'k}
    + (\partial_2 H_{ijk})_l q_{l'k}
    + (\partial_3 H_{ijk})_{lm} \Phi_{lk'k}.
\eea

The conditions \re{pcn1}-\re{pcn3} can be written as
\bea
\partial_1 j_i &=& - (\partial_2 s)_j \partial_1 Q_{ij} -
     (\partial_3 s)_{jk} \partial_1 H_{ijk}  \\
(\partial_2 j_j)_i &=&  \partial_1s \delta_{ij} - (\partial_2 s)_k (\partial_2 Q_{ij})_k
    - (\partial_3 s)_{kl} (\partial_2 H_{ijk})_l \\
(\partial_3 j_k)_{ij} &=& - (\partial_2 s)_l (\partial_3 Q_{ij})_{kl}
    - (\partial_3 s)_{lm} (\partial_3 H_{ijk})_{lm}.
\eea

As a consequence the above system of equations can be solved, as
the entropy current is a potential of the field
$(q_i,Q_{ij},H_{ijk})$, therefore it can be conveniently written
as
\be
 j_i(e, q_i, \Phi_{ij}) = \tilde{j}_i(q_i,Q_{ij},H_{ijk}).
\ee

This fact can be expressed also with differential forms, according
to the traditions of thermodynamics
\be
{\rm d} \tilde{j}_i = \partial_1s {\rm d} q_i
    + (\partial_2s)_{j} {\rm d} Q_{ji}
    + (\partial_3s)_{jk} {\rm d} H_{jki}
= \Gamma^1 {\rm d} q_i
    + \Gamma^2_{j} {\rm d} Q_{ji}
    + \Gamma^3_{jk} {\rm d} H_{jki}.
\label{CurrPot}\ee

The derivatives of the entropy current are identical to the
intensives, the derivatives of the entropy functions. However, the
variables are different. This form results in serious restrictions
of the entropic intensives, and the currents $Q_{ij}$ and
$H_{ijk}$ because the mixed second partial derivatives should be
equal in these variables, too. Expressed in the basic variables
the above requirements are rather ugly
\begin{gather}
(\partial_{22}s)_{kj} \partial_1 Q_{ki}
    + (\partial_{23}s)_{lkj} \partial_1 H_{lki}
= \partial_{11}s \delta_{ij}
    + (\partial_{12}s)_{k} (\partial_2 Q_{ik})_j
    +(\partial_{13}s)_{lk} (\partial_2 H_{lki})_j, \label{jcon1}\\
(\partial_{32}s)_{ljk} \partial_1 Q_{li}
    + (\partial_{33}s)_{lmjk} \partial_1 H_{lmi}
= (\partial_{12}s)_{l} (\partial_3 Q_{li})_{jk}
    + (\partial_{13}s)_{lm} (\partial_3 H_{lmi})_{jk},  \label{jcon2}\\
(\partial_{31}s)_{kl} \delta_{ij}
    + (\partial_{32}s)_{mkl} (\partial_2 Q_{im})_j
    + (\partial_{33}s)_{mnkl} (\partial_2 H_{mni})_j =\nonumber\\
= (\partial_{22}s)_{ml} (\partial_3 Q_{mi})_{jk}
    + (\partial_{32}s)_{mnl} (\partial_3 H_{mni})_{jk}.  \label{jcon3}
\end{gather}

The property \re{CurrPot} is an important consequence of the
balance form of the evolution equations. It is independent of the
choice of the basic variables. If one assumes e.g. that the chosen
internal variable is the current of the heat flux, $\Phi_{ij} =
Q_{ij}$, as it is usual in extended thermodynamics, then the
above system of requirements simplifies but does not disappear.

In Rational Extended Thermodynamics  it was shown that the above
result of the phenomenological theory  can be in accordance with
the kinetic theory of gases, at least with a classical formulation
of kinetic physics. A crucial step in the different systems was
the choice of the phenomenological variables (we will see, that
all of the currents cannot be chosen as internal variables without
any further ado) and the use of source terms in the balances.

\subsection{Local state and balance form evolution: isotropy}

The system \re{jcon1}-\re{jcon3} does not have a general solution
for the currents. $Q_{ij}$ and $H_{ijk}$ cannot be determined by
the entropy function in general. Therefore we have lost one of the
basic flavors of irreversible thermodynamics, that the
requirements of the second law can be exploited constructively to
find a the appropriate evolution equations. Now the dissipation
inequality was solved, but the evolution equations cannot be
determined constitutively.

Jou, Lebon, Mongiovi and Peruzza gave some simplifying conditions
to have a solution of the conditions \re{jcon1}-\re{jcon3} on the
phenomenological level \cite{JouAta04a}. They have assumed a
local state, balance form evolution equations and a simpler set of
variables, they introduced only $q_i$ as an additional
variable. Then only condition \re{jcon1} applies in a simplified
form as
$$
(\partial_{22}s)_{kj} \partial_1 Q_{ki}
= \partial_{11}s \delta_{ij}
    + (\partial_{12}s)_{k} (\partial_2 Q_{ik})_j.
$$

Moreover they have assumed {\em isotropic materials}, when all
scalar valued functions, including the entropy, depend only on
$q^2 = q_iq_i$ and the flux of the heat current and the entropy
current can be written as
\bea
Q_{ij} &=& \beta(e,q^2) + \psi(e,q^2)q_iq_j, \\
j_i    &=& \Psi(e,q^2)q_i.
\eea

Now the requirement \re{CurrPot} results in the following system
of equations, as (4.2)-(4.4) in \cite{JouAta04a}
\bea
\partial_e\Psi &=& 2\partial_{q^2} s(\partial_e\beta + \partial_e\psi q^2), \\
\Psi &=& \partial_e s + 2\partial_{q^2}s \psi q^2, \\
\partial_{q^2} \Psi &=& 2\partial_{q^2} s \left(\partial_{q^2} \beta
    + \partial_{q^2}\psi q^2 \right).
\eea

Therefore the entropy current can be written as
\be
j_i = \left(\partial_es + 2\partial_{q^2}s \psi q^2 \right) q_i.
\label{Jouecurr}\ee

After further calculations, considering also the requirement
\re{jcon1}, one can get explicit solutions for the functions
$\beta$ and $\psi$ together with some additional restriction on
the form of the entropy function.

\section{Local state and special entropy current}

A different solution of the dissipation inequality can be given
with the help of the entropy current. As in our previous work
\cite{CiaAta03m} we consider the local entropy of the form
\re{Gyarent}. This general form is motivated by the requirement of
the thermodynamic stability, or, equivalently the requirement of
the concavity of the entropy function on the non-equilibrium part
of the state space (spanned by $q_i$ and $\Phi_{ij}$). Therefore,
$m_{ij}$ and $n_{ijkl}$ are positive definite constitutive
functions. Moreover, let us assume the entropy current takes the
form \re{Nyiecurr} and let us introduce the convenient notations
$\hat{m}, \hat{n}, \tilde{n}, \tilde{m}$ as follows
\bea
(\partial_2s)_i &=& -m_{ij}q_j -
    \frac{1}{2} (\partial_2 m_{lj})_i q_l q_j -
    \frac{1}{2} (\partial_2 n_{rjkl})_i \Phi_{rj} \Phi_{kl} \nonumber\\
&=& - \hat{m}_{ij} q_j - (\partial_2\tilde{n}_{jk})_i \Phi_{jk}, \label{mnl}\\
(\partial_3 s)_{ij} &=&
    -\frac{1}{2} (\partial_3 m_{lk})_{ij} q_l q_k -
    n_{ijkl}\Phi_{kl} -
    \frac{1}{2} (\partial_3 n_{lkrs})_{ij} \Phi_{lk} \Phi_{rs} \nonumber\\
&=& -(\partial_3 \tilde{m}_k)_{ij} q_k - \hat{n}_{ijkl}\Phi_{kl} \label{nnl}
\eea
If $m_{ij}$ and $n_{ijkl}$ are constant, then $\hat{m}_{ij} =
m_{ij}$, $\hat{n}_{ijkl} = {n}_{ijkl}$, $(\partial_3
\tilde{m}_k)_{ij} = 0$ and $(\partial_2\tilde{n}_{jk})_i = 0$  .

Let us emphasize again that the entropy, written in the Gyarmati
form \re{Gyarent} and the entropy current, written in
\re{Nyiecurr} are only convenient notations as long as the
corresponding inductivities and current multipliers are general
constitutive functions.

With the \re{Nyiecurr} form of the entropy current and using the
notations of \re{mnl} and \re{nnl} the dissipation inequality
\re{Dis1ls} in local state can be written in the following form:
\bea
&& [A_{ji'j} - \hat{m}_{ij} g_j -
    (\partial_3 m)_{kji} f_{ik} ]q_i +
    [A_{ij} - \partial_1s \delta_{ij} ] q_{j'i} + \nonumber\\
&& [B_{kij'k} - (\partial_2 n)_{kji} g_k -
    \hat{n}_{ijkl} f_{kl} ] \Phi_{ij} +
    B_{ijk} \Phi_{jk'i} \geq 0.
\label{dislocnyi}
\eea

Seemingly the system is a normal force-current system, because the
coefficients of the thermodynamic forces $q_i, q_{i,j}, \Phi_{ij},
\Phi_{ijk}$ all contain undetermined constitutive quantities
$A_{ij}, g_i, f_{ij}$, $B_{ijk}$ respectively. However, let us
observe that in local state $g_i$ and $f_{ij}$ are nonlocal but
all other terms are local in the above inequality. This fact
introduces degeneracy since, although the coefficients of the
derivatives $q_{i'j}$ and $\Phi_{ij'k}$ cannot disappear, their
possible couplings  are rather reduced, e.g. $B_{ijk}$ is local,
therefore cannot depend on its own force $\Phi_{ij'k}$.

In this degenerate case the solution of the dissipation inequality
is not straightforward.

Fortunately we can avoid the treatment of degeneracy, e.g. by
assuming that  $m_{ij}$ and $ n_{ijkl}$ depend only on the
internal energy and introducing the  form \re{Verecurr} of the
entropy current with the assumptions $A_{ij} =
\partial_1 s \delta_{ij}$ and $B_{ijk} = (\partial_2s)_k \delta_{ij}$.
In this particular case the dissipation inequality reduces to
\bea \left((\partial_1 s)_{'i} - {m}_{ij} g_j -
    m_{ki} \Phi_{jk'j} \right)q_i +
    \left(-m_{jl} q_{l'i} - {n}_{ijkl} f_{kl}\right) \Phi_{ij}  \geq 0.
\label{dislocver}
\eea
This is a force-current system, with the following Onsagerian
solution
\bea
- {m}_{ij} (\dot{q}_j + \Phi_{kj'k}) + (\partial_1
s)_{'i}  &=&
    L^{11}_{ij} q_j + L^{12}_{ijk}\Phi_{jk}, \label{VerO1}\\
- {n}_{ijkl} \dot{\Phi}_{kl} -m_{jl} q_{l'i}   &=&
    L^{21}_{ijk} q_k + L^{22}_{ijkk}\Phi_{kl}.
\label{VerO2}\eea

The system above conserves the structure already obtained in
\cite{CiaAta03m} in the case of local state. One should emphasize
the central role of the invertibility of the matrices ${m}_{ij}$
and ${n}_{ijkl}$ in order to obtain the balance form. Such a
property is not trivial since there exist real materials for which
it is not guaranteed. A classical example is given in
\cite{Mei73a1}, where an electric circuit described by dynamic
variables is considered.

As it was observed in \cite{CiaAta03m} the second equation is a
phenomenological closure of the system \re{balinte} and
\re{VerO1}. It is remarkable that we have recovered the usual
phenomenological structure of Extended Irreversible Thermodynamics
keeping the entropy current of the form \re{Verecurr}. The same is
not true for the constitutive equation \re{Nyiecurr} which, in the
case of first order nonlocality, seems to be much too general.
However, the compatibility of the entropy current of Verh\'as
\re{Verecurr} and the requirements of the balance form
\re{CurrPot} is valid only with the restriction that the
conductivities depend only on the internal energy. The mentioned
solution of Jou et. al. \re{Jouecurr} clearly does not have the
Verh\'as form and indicates the necessity of a more general
treatment.

In the next section we study the same problem in the presence of
second order nonlocality.

\section{Second order nonlocality - solution in local state}

Now we will extend our investigations to consider second order
nonlocalities. However, for the sake of simplicity we will
investigate only the case of local entropy, i.e. systems in local
state. The basic state is spanned by the variables $(e,q_i,
\Phi_{ij})$ as previously. However, the constitutive space
contains the second order space derivatives and is spanned by
$(e,q_i,\Phi_{ij},e_{'i},q_{i'j},\Phi_{ij'k},e_{'ij}, q_{i'jk},
\Phi_{ij'kl})$. Therefore, in the exploitation of the entropy
inequality \re{balent} we should consider as constraints the
evolution equations \re{balinte}, \re{ev1} and \re{ev2}, their
first prolongations \re{dbalinte}, \re{dev1} and \re{dev2} and
also their second prolongations as follows
\bea
\dot{e}_{'ij} + q_{j'jik} &=& 0, \label{d2balinte}\\
\dot{q}_{i'jk}  - g_{i,jk} &=& 0,    \label{d2ev1}\\
\dot{\Phi}_{ij'kl} - f_{ij'kl} &=& 0.\label{d2ev2}
\eea

Our simplifying condition of local entropy can be written as
$$
s = s(e,q_i,\Phi_{ij}).
$$

Let us introduce again the Lagrange-Farkas multipliers $\Gamma^1$,
$\Gamma^2_i$,  $\Gamma^3_{ij}$, $\Gamma^4_{i}$, $\Gamma^5_{ij}$ and
$\Gamma^6_{ijk}$ for the evolution equations \re{balinte}, \re{ev1},
\re{ev2} and their prolonged forms \re{dbalinte}, \re{dev1} and
\re{dev2} respectively. The multipliers $\Gamma^7_{ij}$,
$\Gamma^8_{ijk}$ and $\Gamma^9_{ijkl}$ stand for the second
prolongations \re{d2balinte}, \re{d2ev1} and \re{d2ev2}
respectively. The Liu procedure gives
\bean
\partial_1s\dot{e} &+&
    (\partial_2s)_i\dot{q}_{i} +
    (\partial_3s)_{ij}\dot{\Phi}_{ij} +\\
+(\partial_1 j_i)e_{'i} &+&
    (\partial_2 j_i)_j q_{j'i} +
    (\partial_3 j_i)_{jk} \Phi_{jk'i} +
    (\partial_4 j_i)_j e_{'ij} +
    (\partial_5 j_i)_{jk} q_{j'ki} +\\
+(\partial_6 j_i)_{jkl} \Phi_{jk'li} &+&
    (\partial_7 j_i)_{jk} e_{'ijk} +
    (\partial_8 j_i)_{jkl} q_{j'kli} +
    (\partial_9 j_i)_{jklm} \Phi_{jk'lmi} - \\
-\Gamma^1\left(\dot{e} + q_{i'i}\right) &-&
    \Gamma^2_i \left(\dot{q}_i  - g_i\right) -
    \Gamma^3_{ij} \left(\dot{\Phi}_{ij}  - f_{ij}\right) -
    \Gamma^4_i (\dot{e}_{'i} + q_{j'ji} ) -\\
-\Gamma^5_{ij} [\dot{q}_{i'j} &-&
    (\partial_1 g_i)e_{'j} -
    (\partial_2 g_i)_k q_{k'j} -
    (\partial_3 g_i)_{kl} \Phi_{kl'j} -
    (\partial_4 g_i)_k e_{'kj} - \\
&-& (\partial_5 g_i)_{kl} q_{k'lj} -
    (\partial_6 g_i)_{klm} \Phi_{kl'mj} -
    (\partial_7 g_i)_{kl} e_{'klj} -\\
&-& (\partial_8 g_i)_{klm} q_{k'lmj} -
    (\partial_9 g_i)_{klmn} \Phi_{kl'mnj}] \\
- \Gamma^6_{ijk} [\dot{\Phi}_{ij'k}  &-&
    (\partial_1 f_{ij})e_{'k} -
    (\partial_2 f_{ij})_l q_{l'k} -
    (\partial_3 f_{ij})_{lm} \Phi_{lm'k} -
    (\partial_4 f_{ij})_l e_{'lk} - \\
&-& (\partial_5 f_{ij})_{lm} q_{l'mk} -
    (\partial_6 f_{ij})_{lmn} \Phi_{lm'nk} -
    (\partial_7 f_{ij})_{lm} e_{'lmk} -\\
&-& (\partial_8 f_{ij})_{lmn} q_{l'mnk} -
    (\partial_9 f_{ij})_{lmno} \Phi_{lm'nok}] \\
-   \Gamma^7_{ij}\left(\dot{e}_{ij} + q_{k'kij}\right) &-&
    \Gamma^8_{ijk} \left(\dot{q}_{i'jk}  - g_{i'jk}\right) -
    \Gamma^9_{ijkl} \left(\dot{\Phi}_{ij'kl}  - f_{ij'kl}\right)
\geq 0. \eean

Here $\partial_n$, $n = 1,2,3,4,5,6,7,8,9$ denotes the partial
derivatives of the constitutive functions according to the
variables of the constitutive space
$(e,q_i,\Phi_{ij},e_{'i},q_{i'j}$, $\Phi_{ij'k},e_{'ij}, q_{i'jk},
\Phi_{ij'kl})$ respectively. The first set of Liu equations
defines the Lagrange-Farkas multipliers as the derivatives of the
entropy and gives that the last six multipliers are zero, due to
the local state.
\bea
\partial_1s &=& \Gamma^1, \label{L21} \\
(\partial_2s)_{i} &=& \Gamma^2_{i}, \label{L22} \\
(\partial_3s)_{ij} &=& \Gamma^3_{ij}, \label{L23} \\
\Gamma^4_{i} &=& 0, \label{L24} \\
\Gamma^5_{ij} &=& 0, \label{L25} \\
\Gamma^6_{ijk} &=& 0, \label{L26}\\
\Gamma^7_{ij} &=& 0, \label{L27} \\
\Gamma^8_{ijk} &=& 0, \label{L28} \\
\Gamma^9_{ijkl} &=& 0. \label{L29}\eea

Considering \re{L24}-\re{L29} the second set of Liu equations are
also simple,
\bea (\partial_7 j_i)_{jk} =0, \label{L210} \\
(\partial_8 j_i)_{jkl} = 0, \label{L211} \\
(\partial_9 j_i)_{jklm} = 0. \label{L212} \eea

Therefore the entropy current presents only first order
nonlocalities
$$
j_i= j_i(e,q_i, \Phi_{ij},e_{'i},q_{i'j},\Phi_{ij'k}).$$

Considering the above conditions the dissipation inequality can be
written exactly in the same form as it was in case of first order
nonlocal constitutive space \re{Dis1ls}:
\bea
(j_i)_{'i} - \partial_1 s q_{i'i} +
    (\partial_2s)_i g_i + (\partial_3 s)_{ij} f_{ij}
    \geq 0.
\label{Dis2ls}\eea

On the other hand now  the constitutive quantities are higher
orderly nonlocal, the entropy current is a first order nonlocal
quantity and ${g_i}$ and $f_{ij}$ are second order nonlocal
quantities. Let us introduce the same assumptions on the form of
the entropy and of the entropy current as previously with assuming
\re{Gyarent} and \re{Nyiecurr} with the notations \re{mnl} and
\re{nnl} but letting the current multipliers $A_{ij}$ and
$B_{ijk}$ to contain first order nonlocalities. The dissipation
inequality can be written in the same form as above
\bea &&
[A_{ji'j} - \hat{m}_{ij} g_j -
    (\partial_3 m)_{kji} f_{ik} ]q_i +
    [A_{ij} - \partial_1s \delta_{ij} ] q_{j'i} + \nonumber\\
&& [B_{kij'k} - (\partial_2 n)_{kji} g_k -
    \hat{n}_{ijkl} f_{kl} ] \Phi_{ij} +
    B_{ijk} \Phi_{jk'i} \geq 0.
\eea

However, in this case it is a non-degenerate force-current system,
due to the extension of the constitutive state space. All additive
terms contain unknown functions. Therefore one can give an
Onsagerian solution as follows
\bea
-\hat{m}_{ij}g_j -(\partial_3 m)_{ijk}f_{jk} +
    A_{ji'j} &=& \nonumber\\
 &&   L^{11}_{ij} q_j + L^{12}_{ijk}q_{j'k} +
    L^{13}_{ijk}\Phi_{jk} + L^{14}_{ijkl}\Phi_{jk'l}, \label{NyiO1}\\
A_{ij} - \partial_1s \delta_{ij}  &=&
    L^{21}_{ijk} q_k + L^{22}_{ijkl}q_{k'l} +
    L^{23}_{ijkl}\Phi_{kl} + L^{24}_{ijklm}\Phi_{kl'm}, \label{NyiO2}\\
- (\partial_2 n)_{kji} g_k - \hat{n}_{ijkl} f_{kl} + B_{kij'k} &=&
\nonumber\\
   && L^{31}_{ijk} q_k + L^{32}_{ijkl}q_{k'l} +
    L^{33}_{ijkl}\Phi_{kl} + L^{34}_{ijklm}\Phi_{kl'm}, \label{NyiO3}\\
    B_{ijk} &=&
    L^{41}_{ijkl} q_l + L^{42}_{ijklm}q_{l'm} +
    L^{43}_{ijklm}\Phi_{lm} + L^{44}_{ijklmn}\Phi_{lm'n}. \label{NyiO4}
\eea

Here $L^{11}, L^{12}, L^{13}, L^{14}, L^{21}, L^{22}, L^{23},
L^{24},L^{31}, L^{32}, L^{33}, L^{34},L^{41}, L^{42}, L^{43},
L^{44}$ are constitutive functions with the suitable definiteness
restrictions. Moreover \re{NyiO1} and \re{NyiO3} are the
candidates of balances under suitable restrictions. It is worth
noticing, that the current multipliers $A_{ij}$ and $B_{ijk}$ are
given explicitly. Therefore, they can be easily eliminated from
the above system substituting \re{NyiO2} into \re{NyiO1} and
\re{NyiO4} into \re{NyiO3}. The resulting set of equations is
closed and contains first and second order derivatives of the
basic state.
\bea
-\hat{m}_{ij}\dot{q}_j
&-& (\partial_3m)_{ijk}\dot{\Phi}_{jk} + \nonumber\\
&+& [\partial_1s \delta_{ij} + L^{21}_{ijk} q_k + L^{22}_{ijkl}q_{k'l} +
    L^{23}_{ijkl}\Phi_{kl} + L^{24}_{ijklm}\Phi_{kl'm}]_{'j} =
    \nonumber\\&=&
    L^{11}_{ij} q_j + L^{12}_{ijk}q_{j'k} +
    L^{13}_{ijk}\Phi_{jk} + L^{14}_{ijkl}\Phi_{jk'l}, \label{NyiBal1}\\
- (\partial_2 n)_{kji} \dot{q}_k &-& \hat{n}_{ijkl} \dot{\Phi}_{kl} + \nonumber\\
&+& [L^{41}_{ijkl} q_l + L^{42}_{ijklm}q_{l'm} +
    L^{43}_{ijklm}\Phi_{lm} + L^{44}_{ijklmn}\Phi_{lm'n}]_{'k}=
    \nonumber \\ &=&
    L^{31}_{ijk} q_k + L^{32}_{ijkl}q_{k'l} +
    L^{33}_{ijkl}\Phi_{kl} + L^{34}_{ijklm}\Phi_{kl'm},\label{NyiBal2}
\eea

As one can see, the evolution equations are rather general. They
are more general that those given with the help of the entropy
currents \re{Verecurr}. On the other hand the entropy current
\re{Jouecurr} is also a special case of the general \re{Nyiecurr}.
However, the compatibility to the current potential structure,
expressed by \re{CurrPot}, cannot be expressed explicitly in
general. \re{Nyiecurr} is definitely more general regarding the
nonlocality, moreover, \re{Jouecurr} proves, that can be
compatible with the potential structure in special cases. However,
it definitely restricts the functional form of the current
multipliers $A_{ij}$ and  $B_{ijk}$.

Moreover, the potential structure is unavoidable requiring
locality and balance form evolution equations. If ${m}_{ij}$ and
${n}_{ijkl}$ are constant a balance structure similar to that
obtained in \cite{CiaAta03m} is recovered. This fact represents
the main effect of the enlargement of the state space since the
balance structure is compatible with more general entropy fluxes.

\section{Nonlocal second sound: the Guyer-Krumhansl equation}

Thermal wave propagation, sometime referred to as second sound, is
a low temperature phenomenon which can be observed, for instance,
in dielectric crystals such as Sodium Fluoride (NaF) and Bismuth
(Bi) \cite{JacWal71a,NarDyn72a,DreStr93a}. It requires an
extension of the classical Fourier's theory in order to remove the
paradox of infinite speed of propagation of thermal disturbances
\cite{Cat48a}. Phonon gas hydrodynamics \cite{Pei55b,Rei73b}
supplies a satisfactory explanation of heat transport at low
temperature. Phonons are quasi-particles which obey the
Bose-Einstein statistics. In a solid crystal  they form a rarefied
gas whose kinetic equation can be derived similarly to that of an
ordinary gas. Phonons may interact among themselves and with
lattice imperfections through two different types of
processes: \\
i) Normal-(N) processes, that conserve the phonon momentum;\\
ii) Resistive-(R) processes, in which the phonon momentum is not
conserved. \\
The frequencies $\nu_{N}$ and $\nu_{R}$ of normal and resistive
processes determine the characteristic relaxation times
$\tau_{N}=\frac{1}{\nu_{N}}$ and $\tau_{R}=\frac{1}{\nu_{R}}$.
Diffusive processes take over when there are many more R-processes
than N-processes. If instead there are only few R-processes and
many more N-processes, then a wave like energy transport may
occur. \\
Such a phenomenology is satisfactorily described by  the
Guyer-Krumhansl equation \cite{GuyKru66a1}, eq. (57) (see also
\cite{GuyKru64a,GuyKru66a2}).
\be
\dot{q_i} + \frac{1}{\tau_R}q_i + \frac{1}{3}c^2 c_V T_{'i} =
    \frac{1}{5} \tau_N c^2 \left(q_{i'jj} + 2 q_{j\;'ji} \right),
\label{Guyer}\ee
where $T=\hat{T}-T_0$ is the temperature variation ($T$ is the
temperature and  $T_0$ is the average temperature). $c_V$ is the
specific heat and $c$ means the Debye phonons velocity. Such an
equation, which generalizes the Maxwell-Cattaneo-Vernotte equation
\cite{Cat48a} \be \tau_{R}\dot{q_i} + q_i = -\tau_{R}
\frac{1}{3}c^2 c_V T_{'i}, \label{catt}\ee
was the first in the literature to include both relaxation times.
It can be obtained by the linearized Boltzman equation for phonons
in the Debye approximation, if one maintains terms $O(\tau_N)$
\cite{GuyKru66a1}. The material coefficients $\tau_N, \tau_R$ and
$c_V$ are all depend on the temperature. According to experimental
observations \cite{AckGuy68a} one can get
$$
c_V = a T^3, \qquad \tau_R = d\; e^{-\alpha/T}, \qquad \tau_N = b
T^{-m}.
$$

Here $a, b, d, \alpha$ are constant coefficients and $m \in \{3,
4, 5\}$ depending on the material. Let us remark that the last
function, a temperature dependent $\tau_N$ is clearly contradicts
to the assumptions made by Guyer and Krumhansl \cite{GuyKru66a1}
p771, and did not consider, that \re{Guyer} is a linearized
equation.

In Extended Thermodynamics the Guyer-Krumhansl equation can be
recovered in a 4-field theory provided one assumes weakly nonlocal
constitutive equations for the internal energy $e$ and for the
momentum flux \cite{DreStr93a}.

However, such an approach seems to be questionable because:\\
i) a nonlocal constitutive space is in contrast with the basic
assumptions of Extended Thermodynamics.\\
ii) a nonlocal internal energy  does not assure that the specific
heat $c_V = \frac{\partial e}{\partial T}$ is positive.\\
iii) a nonlocal internal energy would modify also the energy balance.

The last observation was pointed also by Dreyer and Struchtrup and
they suggested to consider higher order moments
\cite{JouAta92b,MulRug98b}. The other important question is the
temperature dependency of the coefficients. A phenomenological
theory cannot predict the exact form to the constitutive
functions, but gives restrictions and interrelations. These
restrictions on the temperature dependencies are frequently
treated rather loosely to get the compatibility with the {\em
linearized} kinetic theory \cite{GuyKru66a1,JouAta92b}.

These problems do not arise in the present theory. In fact, we can
obtain the equation \re{Guyer} by considering a 4-field model with
second order nonlocality, namely the balance equation \re{NyiBal1}
with $\Phi_{ij} = 0$ and $m_{ij}$ depending only on the internal
energy. It yields
\be
 -m_{ij}\dot{q}_j + \Big[\partial_1s \delta_{ij} + L^{21}_{ijk}
 q_{k} + L^{22}_{ijkl}q_{k'l}\Big]_{'j} =
    L^{11}_{ij} q_j + L^{12}_{ijk}q_{j'k}. \label{NyiBal12}\\
\ee
It is convenient to choose as equilibrium thermodynamic variable
the absolute temperature $T_a$ instead of the internal energy $e$.
Let us assume the constitutive equations $L^{21}_{ijk}=
L^{12}_{ijk} = 0$.
we may exploit the thermodynamic relation $\frac{1}{T_a} =
\frac{\partial s}{\partial e}$ and introduce the temperature
perturbation $T$ by $T_a = T_0 + T$, where $T_0$ is a background
(average) temperature. Now, equation \re{NyiBal12} reduces to
\be m_{ij}\dot{q}_j + L^{11}_{ij} q_j + \frac{1}{T^2}T_{'i} =
L^{22}_{ijkl}q_{k'lj}. \label{NyiBal123}\\
\ee
Here ${\bf m}$, ${\bf L}^{11}$ and ${\bf L}^{22}$ can depend on
the background temperature $T_0$. Finally, equation \re{Guyer} is
easily obtained by \re{NyiBal123} under the further constitutive
assumptions
\be \label{const2} m_{ij} =\frac{3}{c^2 a T^5} \delta_{ij}, \ee
\be \label{const3} L^{11}_{ij} =\frac{3}{c^2 a T^5}
e^{\frac{\alpha}{T}} \delta_{ij}, \ee
\be \label{const4} L^{22}_{ijkl} =\frac{3 b}{5 a T^{5+m}}
\delta_{ij}\delta_{kl}.\ee
That way, the Guyer-Krumhansl theory may be obtained in a classical
macroscopic framework. Let us remark, that the previous
phenomenological linearization is only a last step in the
calculation based on the linearization of the Boltzmann equation
(eq. (57) to eq. (59) in \cite{GuyKru66a1}). A similar result has
been obtained in \cite{Van01a2}.

\section{Conclusions}

We have shown that Classical Irreversible Thermodynamics
supplemented with dynamic degrees of freedom is consistent with the
idea of higher order fluxes as independent thermodynamic variables.
In such a framework the balance form of the evolution equations can
be obtained under suitable constitutive assumptions. The locality of
the entropy density, i.e. the local state assumption, plays a
central role. Moreover, the nonlocality of the constitutive
functions results in a wide class of materials including the
classical Cattaneo's and Guyer-Krumhansl heat conductors, which
normally are derived from kinetic theory. Let us observe that the
model above encompasses all extended thermodynamic models, since the
balance structure represents a very particular form of nonlocal
evolution equations \re{ev1} and \re{ev2}.

The figure below shows the connections between the constitutive
assumptions and the thermodynamic restrictions, as far as the
locality and nonlocality are concerned. The last column refers to
the corresponding section of the paper. $L_A$ denotes a locality
assumption. The double head arrows denote specific interrelations,
e.g. the black double head arrows in the first row denote equations
\re{L7}-\re{L9}, that give conditions between the entropy and
entropy current functions and evolution equations. The fifth row
refers to Rational Extended Thermodynamics and the sixth row to the
specific solution of the potential restriction given by Jou, Lebon,
Mongiovi and Peruzza. As in Extended Thermodynamics the dissipations
inequality is fulfilled as an equality, we cannot conclude anything
on the dissipative constitutive functions (that is why there are no
white arrows in these rows). Let us observe, that assumptions on the
form of the entropy current makes possible to build all requirements
of the Second Law into the evolution equations in general (last two
rows). In this case the Second Law become a material property,
satisfied independently of the initial conditions and the resulted
constitutive functions depend only on the material.

It is worth noticing that the potential form of the entropy current
and the balance structure is compatible with the general form of the
entropy current, such as that proposed by Ny\'\i{}ri, which yields
\re{Verecurr} when $A_{ij} = \partial_1 s \delta_{ij}$ and $B_{ijk}
= (\partial_2s)_k \delta_{ij}$. In this case the balance structure
results in restrictions for the current multipliers from the
potential requirement \re{CurrPot}.

We obtained closure relations both with local and nonlocal entropy
current and in the first case the closure for the highest order flux
was an ordinary differential equation. The obtained thermodynamic
closure of the hierarchical structure stresses some deeper relations
between the thermodynamic and the more detailed kinetic structure,
similar that was recognized e.g. in \cite{Dre87a}.

We reinspected the thermal wave propagation at low temperature and
proved that the Guyer-Krumhansl theory of second sound can be
derived in the framework of macroscopic nonlocal irreversible
thermodynamics.

\section{Acknowledgements}

This research was supported by Progetto COFIN 2002: Modelli
matematici per la scienza dei materiali, and by the grants OTKA
T034715 and T034603. The University of Basilicata is acknowledged as
well.

\vfill\pagebreak
\begin{figure}
\centering
\includegraphics[height=22cm]{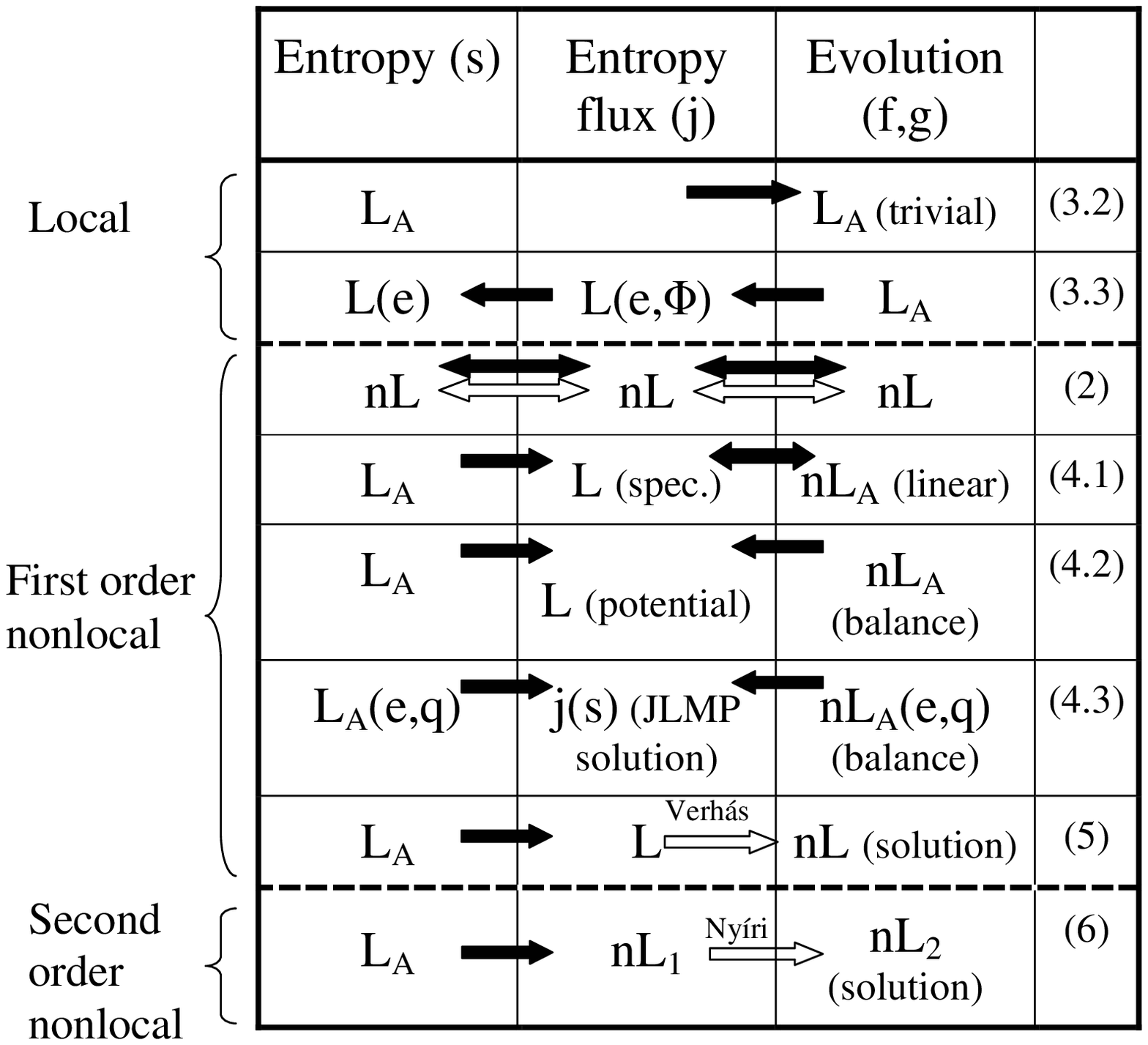}
\caption{Notations: L - local, nL - nonlocal, L(e) - local in the
variable e, $L_A$ - assumed locality, $L_i$ - i-th order nonlocal,
Black arrow - consequence of the Liu equations, White arrow -
consequence of the dissipation inequality} \label{Fig1}
\end{figure}

\bibliographystyle{unsrt}

\vskip 1cm

V. A. Cimmelli, \\Department of Mathematics, \\University of
Basilicata,\\
Campus Macchia Romana, 85100 Potenza, Italy \\
E-mail:cimmelli@unibas.it\\

P. V\'an, \\Department of Chemical Physics, \\Budapest University of
Technology and Economics, \\Budafoki \'ut 8, 1521 Budapest, Hungary \\
E-mail:vpet@phyndi.fke.bme.hu \\

\end{document}